\documentclass{article}
\usepackage[preprint]{spconfa4}
\usepackage[linesnumbered,algoruled]{algorithm2e}
\usepackage{amsmath,amssymb,amsfonts,url}
\usepackage{multirow}
\usepackage{graphicx,color}
\usepackage[dvipsnames]{xcolor}
\title{Harmonic-Percussive Source Separation with\\ Deep Neural Networks and Phase Recovery}
\name{Konstantinos Drossos$^{*\dagger}$, Paul Magron$^{*\dagger}$, Stylianos Ioannis Mimilakis$^{*\ddagger}$, and Tuomas Virtanen$^{\dagger}$\thanks{$^{*}$Equally contributing authors.}}
\address{$^{\dagger}$Laboratory of Signal Processing, Tampere University of Technology, Tampere, Finland \\
\texttt{\{firstname.lastname\}@tut.fi}\\
$^{\ddagger}$Fraunhofer IDMT, Ilmenau, Germany\\
\texttt{mis@idmt.fraunhofer.de}}
\begin{document}
\ninept
\maketitle
\begin{abstract}
Harmonic/percussive source separation (HPSS) consists in separating the pitched instruments from the percussive parts in a music mixture. In this paper, we propose to apply the recently introduced Masker-Denoiser with twin networks (MaD TwinNet) system to this task. MaD TwinNet is a deep learning architecture that has reached state-of-the-art results in monaural singing voice separation. Herein, we propose to apply it to HPSS by using it to estimate the magnitude spectrogram of the percussive source. Then, we retrieve the complex-valued short-time Fourier transform of the sources by means of a phase recovery algorithm, which minimizes the reconstruction error and enforces the phase of the harmonic part to follow a sinusoidal phase model. Experiments conducted on realistic music mixtures show that this novel separation system outperforms the previous state-of-the art kernel additive model approach.
\end{abstract}
\section{Introduction}
\label{sec:intro}
Audio source separation~\cite{Comon2010} consists in extracting the underlying \textit{sources} that add up to form the observed audio \textit{mixture}. Harmonic/percussive source separation (HPSS)~\cite{Ono2008} is a particular case of the audio source separation task which aims at segregating the percussive sounds (such as drum hits) from the pitched instrument components (such as guitar and piano notes). HPSS is a useful preprocessing tool for many applications spanning from music information retrieval to digital audio effects. For instance, the percussive components of a music mixture can be used to estimate the beat of a music recording~\cite{Gkiokas2012}. 
On the other hand, the performance of audio effects, such as time-stretching, can be significantly improved by manipulating the harmonic components only~\cite{Driedger2014}.

\begin{figure}[t]
	\hspace{-2em}
    \centering
	\includegraphics[width=.95\columnwidth]{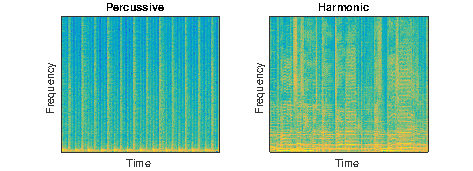}
	\caption{Percussive (left) and harmonic (right) signals spectrograms.}
    \vspace{-1em}
	\label{fig:hp_spec}
\end{figure}

HPSS techniques commonly act on a time-frequency (TF) representation of the data, such as the short-time Fourier transform (STFT). An example of STFT magnitude is illustrated in Fig.~\ref{fig:hp_spec}, in which the structure of music instruments is more prominent: the percussive sounds are usually localized in time and spread across frequencies (vertical lines), while harmonic components are sparse in frequency and are activated over time (horizontal lines).

Traditional methods for HPSS consists in filtering the data in the TF domain in order to exploit this particular structure of percussive and harmonic sound events. Median filtering~\cite{FitzGerald2010} operates in both directions (along frequencies and time) of mixture magnitude spectral representations to segregate vertical and horizontal components. It was later improved by using several different filters and inheriting the additivity constrain for the sources, resulting into the kernel additive model (KAM)~\cite{FitzGerald2013,Liutkus2014}. These particular shapes in the TF domain are also used in an optimization framework as regularizations to estimate the sources~\cite{Ono2008,Tachibana2014}. Similarly in~\cite{Cano2014}, the phase information of the mixture signal is used to exploit the structure of harmonic and percussive instruments for refining the TF filtering process. Alternatively, models based on non-negative matrix factorization (NMF)~\cite{Lee1999} have been applied to this task. Specific NMF methods for the task of HPSS include the use of constraints such as sparsity of percussive sources along the direction of time~\cite{Canadas2014}, structured factorization models that take into account the quasi-stationarity of harmonic sources~\cite{Laroche2015} and extensions of NMF that account for the non-stationarity of percussive components~\cite{Smaragdis2004,Laroche2017}.

However, state-of-the-art results for source separation are obtained with deep learning methods~\cite{Nugraha2016,Takahashi2017}, which learn the model from the given data. They have shown particularly successful for the task of singing voice separation~\cite{Mimilakis2017,Mimilakis2018,Drossos2018}. Recently, a DNN-based HPSS method has been introduced~\cite{Lim2017} and is based on learning a set of convolution kernels that perform the separation. That method has shown significant improvement over traditional approaches.

In this paper, we propose a novel DNN-based method for HPSS. The method is based on the Masker-Denoiser architecture~\cite{Mimilakis2018} regularized using the twin networks (MaD TwinNet) as proposed in~\cite{Drossos2018}. This DNN topology was initially designed for monaural singing voice separation, achieving state-of-the-art results. Given a set of observed mixture magnitude spectra, MaD TwinNet can generate source dependent TF masks that are applied to the mixture magnitude spectra, yielding the magnitude spectrogram of the target source. In the context of this work, we use MaD TwinNet for the task of HPSS, by training it in a supervised scenario to yield estimates of the percussive magnitude spectrogram.

Since phase information plays an important role not only in the source identification process~\cite{Cano2014} but also for the source reconstruction~\cite{Magron2018}, we propose to apply a recently-introduced phase recovery algorithm that exploits a sinusoidal model~\cite{Magron2018} and combine it with the magnitude estimates obtained with MaD TwinNet. Given that percussive sounds are not well represented with mixtures of sinusoids, we adapt this algorithm to the specific task of HPSS, where the sinusoidal model is only promoted for the harmonic part. We test the proposed technique on professionally produced music recordings, that are used in the signal separation evaluation campaign (SiSEC)~\cite{Liutkus2017}. 
The approach based on MaD TwinNet combined with the phase recovery algorithm proposed in~\cite{Magron2018} yields results that surpass the KAM algorithm~\cite{Liutkus2017} by a large margin.

The rest of this paper is structured as follows. Section~\ref{sec:dnn} presents the MaD TwinNet system which predicts the percussive magnitude spectrum. Section~\ref{sec:phase} introduces the phase recovery algorithm. The experimental setup is detailed in Section~\ref{sec:exp_setup}. The results are reported in Section~\ref{sec:exp_results}, and Section \ref{sec:conclu} draws some concluding remarks.
\vspace{-6pt}
\section{MaD TwinNet}\label{sec:dnn}
\begin{figure}[t]
	\centering
	\includegraphics[width=.99\columnwidth, keepaspectratio]{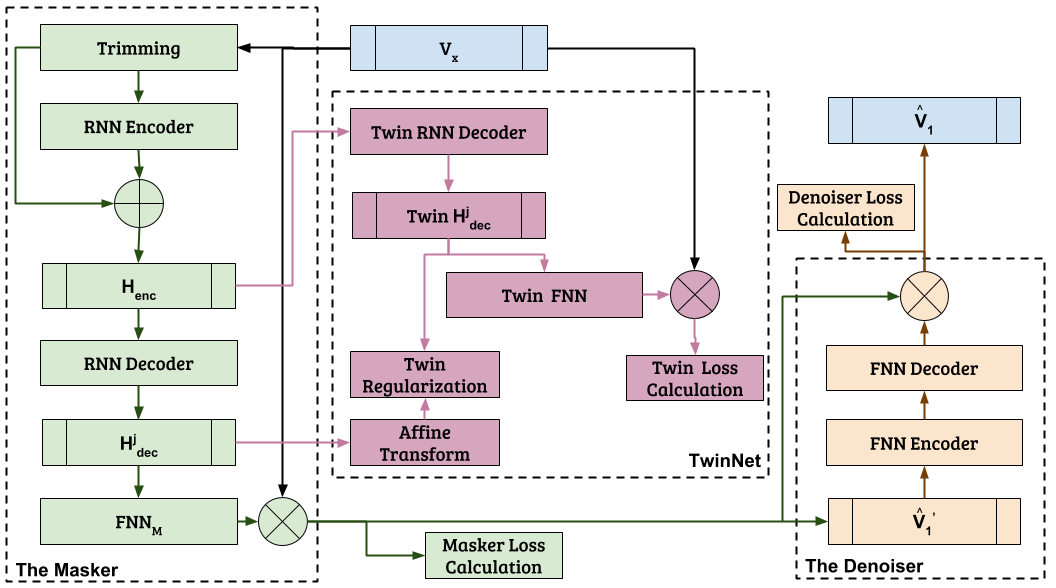}
	\caption{Illustration of the Mad TwinNet system (adapted from~\cite{Drossos2018}). With green color is the Masker, with light brown the Denoiser, and with magenta the TwinNet regularization.}
    \vspace{-1em}
	\label{fig:madtwinnets}
\end{figure}

\vspace{-6pt}
We present here MaD TwinNet, which is used as a core system in our separation framework. Indeed, this deep learning model has shown to be the most up-to-date for singing voice separation~\cite{Drossos2018}, and therefore we propose to use it for an HPSS task. This architecture is a compound system which consists of two components, namely the Masker and the Denoiser, to which is added a regularization based on twin networks (TwinNet)~\cite{Serdyuk2018}, and it is illustrated in Fig.~\ref{fig:madtwinnets}. We briefly explain it hereafter, and more details on it can be found in~\cite{Mimilakis2018,Drossos2018}.

\subsection{The Masker and the Denoiser}
\vspace{-3pt}
The input to the Masker is the magnitude spectrogram of the mixture, $\mathbf{V}_{x}$, and the output of its last layer is a TF mask that is applied to $\mathbf{V}_{x}$. This masking process produces the output of the Masker, which is a first estimate of the percussive magnitude spectrogram, denoted $\hat{\mathbf{V}}_1'$. The latter is used as an input to the Denoiser, whose last layer outputs a TF filter that acts as a denoising filter upon $\hat{\mathbf{V}}_1'$. The output of the denoising process in the Denoiser is the final estimate of the magnitude spectrogram, $\hat{\mathbf{V}}_{1}$. Both the Masker and the Denoiser are based on the denoising auto-encoders framework~\cite{vincent:2010:jml}.
More specifically, the Masker consists of a bi-directional recurrent neural network (RNN) encoder ($\text{RNN}_{\text{enc}}$), that accepts as an input the magnitude spectrogram of the mixture $\mathbf{V}_{x}$ and iterates over the rows of $\mathbf{V}_{x}$. The output of the $\text{RNN}_{\text{enc}}$ is used in a residual connection with the input $\mathbf{V}_{x}$, producing $\mathbf{H}_{\text{enc}}$. $\mathbf{H}_{\text{enc}}$ is used as an input to a forward RNN decoder ($\text{RNN}_{\text{dec}}$), which outputs the hidden states $\mathbf{H}^{1}_{\text{dec}}$. The latter is then given as an input to a sparsifying transform, i.e. a feed-forward neural network ($\text{FNN}_{\text{M}}$) followed by a rectified linear unit (ReLU), in order to produce a TF mask $\mathbf{M}$. This mask, along with the mixture's magnitude spectrogram $\mathbf{V}_{x}$, are multiplied by a skip-filtering connection to produce the first estimate of the targeted magnitude spectrogram:
\begin{equation}
\hat{\mathbf{V}}_1'= \mathbf{M} \odot\mathbf{V}_{x},
\end{equation}
where $\odot$ is the Hadamard product. 

$\hat{\mathbf{V}}_1'$ is expected to contain interferences from other music sources \cite{Mimilakis2017,Mimilakis2018}. Therefore, MaD utilizes another module, the Denoiser, on top of the Masker, which consists of two feed-forward layers denoted $\text{FNN}_{\text{enc}}$ and $\text{FNN}_{\text{dec}}$. $\text{FNN}_{\text{enc}}$ and $\text{FNN}_{\text{dec}}$ implement an encoding and a decoding stage respectively, and each one is followed by a ReLU non-linearity. The output of the $\text{FNN}_{\text{dec}}$ and $\hat{\mathbf{V}}_1'$ are multiplied by a skip-filtering connection, producing the final magnitude spectrogram by the MaD architecture, $\hat{\mathbf{V}}_1$.
\vspace{-3pt}
\subsection{Twin network regularization}
\vspace{-3pt}
Music signals are governed by long term temporal patterns, like melody and rhythm. RNNs may appear as an appropriate tool for accounting for such temporal patterns. However, the learning signal from RNNs can be dominated by local time structures that impede the learning of the longer term temporal patterns of a musical signal~\cite{bengio:1994:tnp, hochreiter:1997:lsm}. There are various approaches that aim at overcoming this issue~\cite{grave:2016:arxiv,gulcehre:2017:arxiv,dieng:2016:arxiv}. The most recent one is the twin networks (TwinNet) regularization, which uses the hidden states of a backward RNN to regularize the hidden states of a forward RNN~\cite{Serdyuk2018}, while both of the RNNs are trained to minimize the same cost. This regularization results in enforcing the forward RNN to take into account the future evolution of the signal (provided by the hidden states of the backward RNN) and thus, make the learning signal for the forward RNN not to be governed by local structures~\cite{Serdyuk2018}. More details can be found in the paper in which TwinNet is introduced~\cite{Serdyuk2018}.

For the current work, TwinNet is used to regularize the hidden states of the $\text{RNN}_{\text{dec}}$ in the Masker. TwinNet is implemented with a backward RNN and a sparsifying transform, replicating (hence the term ``twin") the $\text{RNN}_{\text{dec}}$, the $\text{FNN}_{\text{M}}$, and the ReLU of the Masker. TwinNet is optimized jointly with the MaD, having the same target and cost function as the Masker does. The input to the TwinNet is the $\mathbf{H}_{\text{enc}}$. The regularization of the $\text{RNN}_{\text{dec}}$ using TwinNet is utilized by minimizing the following cost:
\begin{equation}
\mathcal{L}^{\text{twin}} = \sum\limits_{t}||\psi(\mathbf{h}_{\text{dec}_{t}}) - \mathbf{h}_{\text{twin}_{t}}||,
\end{equation}
where $\psi$ is a trainable affine transform, $\mathbf{h}_{\text{twin}_{t}}$ is the hidden state of the backward RNN of the TwinNet and $||.||$ is the Frobenius norm.
\vspace{-6pt}
\section{Phase recovery}\label{sec:phase}
\vspace{-6pt}
Once we have an estimate of the percussive magnitude spectrum $\hat{\mathbf{V}}_1$, we retrieve the harmonic magnitude as $\hat{\mathbf{V}}_2 = \mathbf{V}_x - \hat{\mathbf{V}}_1$. Then, it is necessary to estimate the phase of those sources in order to retrieve estimates of their complex-valued STFT. A baseline approach~\cite{Drossos2018} consists in using the mixture's phase:
\begin{equation}
\forall j \in \{ 1,2\} \text{, } \hat{\mathbf{S}}_j = \hat{\mathbf{V}}_j \odot\ e^{\odot i \angle \mathbf{X}},
\end{equation}
where $\angle .$ denotes the complex argument and $.^{\odot}$ denotes the element-wise matrix power. Retrieving the complex-valued STFTs by using the mixture's phase is justified in TF bins where only one source is active. Indeed, in such a scenario, the mixture is equal to the active source. However, this is not the case in TF bins where sources overlap, which is common in music signals. This motivates the use of improved phase recovery techniques for addressing this issue.

Here, we propose to adapt the phase retrieval algorithm introduced in~\cite{Magron2018} to the specific case of HPSS. This approach aims at minimizing the mixing error:
\begin{equation}
\mathcal{C}(\hat{\mathbf{S}}) = || \mathbf{X}- \hat{\mathbf{S}}_1 - \hat{\mathbf{S}}_2||^2,
\label{eq:mixerror}
\end{equation}
subject to $|\hat{\mathbf{S}}_j| = \hat{\mathbf{V}}_j$ $\forall j$. An iterative scheme is obtained by using the auxiliary function method which provides updates on $\hat{\mathbf{S}}_j$. In a nutshell, it consists in computing the mixing error at one given iteration, distributing this error onto the estimated sources with a gain:
\begin{equation}
\mathbf{G}_j = \frac{\hat{\mathbf{V}}_j^{\odot 2}}{\hat{\mathbf{V}}_1^{\odot 2}+\hat{\mathbf{V}}_2^{\odot 2}},
\label{eq:gain_wien}
\end{equation}
and then normalizing the obtained variables so that their magnitude is equal to the target magnitude values $\hat{\mathbf{V}}_j$.

The key idea of the algorithm is to initialize the phase of the harmonic track estimates $\hat{\mathbf{S}}_2$ with the values provided by the sinusoidal model, which is widely used for representing audio signals~\cite{Krawczyk2014}. This approach consists in modeling the harmonic part as a sum of sinusoids, from which we can explicitly compute the STFT phase. This leads to the following the \textit{phase unwrapping} (PU) equation for the phase of the harmonic part denoted $\boldsymbol{\Phi}_2$:
\begin{equation}
\phi_{2,ft} \approx \phi_{2,ft-1} + 2 \pi l \nu_{ft},
\label{eq:phase_unwrapping}
\end{equation}
where $l$ is the hop size of the STFT and $\nu_{ft}$ is the normalized frequency in channel $f$ and time frame $t$. As in~\cite{Magron2018}, these frequencies are estimated by means of a quadratic interpolated FFT~\cite{Abe2004} on the log-spectra of the harmonic magnitude estimate $\hat{\textbf{V}}_2$. This estimation is performed in each time frame in order to account for frequency variations. Note that the model~\eqref{eq:phase_unwrapping} is valid only under the assumption that at most one sinusoidal component is active per frequency channel, which we will assume to be the case here. For more details about this model, we refer the interested reader to~\cite{Magron2018}.

Therefore, we use this model to initialize the phase of the harmonic component in our procedure. The phase of the percussive track is initialized with the mixture's phase. This results in a fast procedure (mixture's phase information is expected to be close to a local minimum with respect to the true source) and the output estimates benefit from the temporal continuity property of the sinusoidal phase model. This procedure, denoted as PU-HPSS, is summarized in Algorithm~\ref{al:unwrap_mix}, where lower-case letters (e.g., $\hat{v}_{j,ft}$) correspond to entries of matrices denoted in bold capital letters (e.g., $\hat{\mathbf{V}}_j$).
\begin{algorithm}[t]
	\caption{PU-HPSS}
	\label{al:unwrap_mix}
			\KwData{Mixture $\mathbf{X}$, magnitudes $\hat{\mathbf{V}}_j$, gains $\mathbf{G}_j$ according to~\eqref{eq:gain_wien}, and frequencies $\boldsymbol{\nu}$}
			\KwResult{Estimated sources $\hat{\mathbf{S}}_j$}
            \tcc{Initialize first frame with the mixture's phase}
            $\forall j$, $\hat{s}_{j,f0} \leftarrow v_{j,f0} e^{i \angle x_{f0}}$\;
			\For{$t:=1$ \KwTo $T-1$}{
			\tcc{Sinusoidal model only for the harmonic part}
			$\phi_{1,ft} \leftarrow \angle x_{ft}$\;
            $\phi_{2,ft} \leftarrow \angle \hat{s}_{2,ft-1} + 2 \pi l \nu_{ft}$\;
			$\forall j$, $\hat{s}_{j,ft} \leftarrow \hat{v}_{j,ft} e^{i \phi_{j,ft}}$\;
            \tcc{Iterative loop}
			\For{$it:=1$ \KwTo $max \_ iter$}{
			 $y_{j,ft} \leftarrow \hat{s}_{j,ft} + g_{j,ft}(x_{ft}-\sum_j \hat{s}_{j,ft}) $\;
			 $\hat{s}_{j,ft} \leftarrow \hat{v}_{j,ft}\frac{y_{j,ft}}{|y_{j,ft}|}$\;
			}
			}
\end{algorithm}
\vspace{-6pt}
\section{Experimental setup}\label{sec:exp_setup}\vspace{-6pt}
\begin{table*}[t]
	\center
    \caption{Median source separation performance over the DSD100 test dataset. Higher is better.}
	\label{tab:ssep}
	\begin{tabular}{ll|ccc|ccc|ccc}
        &  & \multicolumn{3}{c|}{Percussive} &  \multicolumn{3}{c|}{Harmonic} &  \multicolumn{3}{c}{Average}  \\
		&  & SDR & SIR & SAR  & SDR & SIR & SAR   & SDR & SIR & SAR \\
        \hline
        \multirow{3}{*}{Setting 1}
& KAM & $1.42$  & $0.44$ & $3.76$ & $6.60$  & $6.71$  & $\textbf{17.66}$ & $4.01$  & $3.57$ & $\textbf{10.71}$  \\
& MaD TwinNet + mixture phase   & $3.35$  & $4.65$ & $\textbf{6.10}$ & $\textbf{8.62}$  & $14.22$ & $10.75$ & $\textbf{5.99}$  & $9.44$ & $8.43$ \\
& MaD TwinNet + PU-HPSS     & $\textbf{3.35}$  & $\textbf{4.66}$ & $6.08$ & $8.58$  & $\textbf{14.45}$ & $10.59$ & $5.97$  & $\textbf{9.55}$ & $8.34$ \\
       \hline
       \multirow{3}{*}{Setting 2}
& KAM & $0.98$  & $5.03$ & $-1.17$ & $6.35$  & $6.58$  & $\textbf{18.51}$ & $3.66$  & $5.80$ & $8.67$  \\
& MaD TwinNet + mixture phase   & $\textbf{3.60}$  & $4.73$ & $\textbf{6.07}$  & $\textbf{8.70}$  & $12.84$ & $11.78$ & $\textbf{6.15}$  & $8.79$ & $\textbf{8.92}$ \\
& MaD TwinNet + PU-HPSS     & $3.59$  & $\textbf{4.76}$ & $6.00$  & $8.69$  & $\textbf{13.11}$ & $11.57$ & $6.14$  & $\textbf{8.94}$ & $8.78$ \\
	\end{tabular}
\end{table*}

As audio material, we used the Demixing Secret Dataset (DSD100), a semi-professionally mixed set of music songs used for the source separation evaluation campaign (SiSEC)~\cite{Liutkus2017}. The dataset is split into two sets (training and testing) consisting of 50 music recordings each, sampled at $44100$ Hz. For each recording, four music sources are available: these are the \textit{bass}, \textit{drums}, \textit{vocals}, and \textit{other} tracks. Using that information from each recording, we synthesize a mixture of $J=2$ sources: the percussive source is equal to the \textit{drums} track, and the harmonic source is obtained by summing the other tracks. Those recordings are down-mixed to monaural signals by averaging the two channels available.

For training MaD TwinNet, we use the ground truth STFT magnitude of the percussive source as target and we optimize the parameters of our method to minimize the generalized Kullback-Leibler divergence between the predicted and the ground truth STFT magnitude spectra. The divergence is computed at both the outputs of the Masker and the Denoiser, and then linearly combined as proposed in the original paper of MaD TwinNet~\cite{Drossos2018}. We use the same Mad TwinNet architecture as in~\cite{Drossos2018} and we set the sequence length equal to 60 time frames, and the context information for the encoding stage in the Masker equal to 10 time frames. For computing the mixture and target source magnitudes, two settings are considered. In the first setting, the STFT is computed with a 46 ms long Hamming window, with a padding factor of 2 and a hop size of 9 ms. In the second setting, the STFT is computed with a 92 ms long Hamming window, with no zero-padding and a hop size of 23 ms. The first setting was used in the original MaD TwinNet paper~\cite{Drossos2018} and is meant to yield good quality magnitude estimates. The second setting corresponds to a scenario where the phase recovery algorithm performs better~\cite{Magron2018}. However, this could result in sacrificing the quality of magnitude estimation, thus reducing the overall performance.

We test MaD TwinNet combined with the mixture phase for estimating the complex STFTs, and we also test the proposed PU-HPSS, which uses 50 iterations. As a comparison baseline, we consider the unsupervised KAM method~\cite{Liutkus2014} which is consider as one the most state-of-the-art methods for HPSS. Even though the DNN-based framework in~\cite{Lim2017} would have been an
appropriate comparison reference, we were unfortunately not able to re-implement it.

Source separation performance is measured with the signal-to-distortion, signal-to-interference, and signal-to-artifact ratios (SDR, SIR, and SAR)~\cite{Vincent2006} expressed in dB and calculated with the \texttt{mir\_eval} Python toolbox~\cite{Raffel2014}. Following the setup used in the SiSEC challenge~\cite{Liutkus2017}, the measures are computed on sliding windows of 30 seconds with 15 second overlap.

In the spirit of reproducible research, the code of this experimental study and MaD TwinNet are available online\footnote{\url{https://github.com/magronp/phase-hpss}}\textsuperscript{,}\footnote{\url{https://github.com/dr-costas/mad-twinnet}}.
\vspace{-6pt}
\section{Results}
\label{sec:exp_results}
\vspace{-6pt}
The separation results on the DSD100 test set are presented in Table.~\ref{tab:ssep}. For a subjective evaluation, there is an online demo with the audio results of the paper\footnote{\url{http://arg.cs.tut.fi/demo/hpss-madtwinnet}}. Firstly, we observe that Mad TwinNet approach outperforms KAM in the percussive part in terms of SDR, SIR and SAR, and also outperforms KAM in the harmonic estimates in terms of distortion and interference reduction. However, KAM yields harmonic estimates which contain less artifacts than Mad TwinNet in both settings. Overall, the DNN-based approach yields better separation results on average in terms of SDR and SIR, while KAM reduces artifacts in the first setting compared to MaD TwinNet.

Secondly, we note that the PU-HPSS algorithm reduces the interference compared to using the mixture's phase, even though this is at the cost of a very moderate drop in SAR and SDR. This highlights the potential of such a sinusoidal model-based phase retrieval algorithm for reducing interference in the estimated signals~\cite{Magron2018c}. However, this improvement in SIR is relatively limited (approximately $0.1$ dB on average). This confirms that the full potential of phase recovery algorithms is only revealed when the magnitudes estimated beforehand (here, with MaD TwinNet) are of relatively good quality~\cite{Magron2018}.

A comparison between the two settings shed some light on how to exploit the separation system at its best potential. Setting 2 leads to overall better results in SDR and SIR for MaD TwinNet. This setting also leads to a better SIR for the percussive part, but lower for its harmonic counterpart. On average, this second setting leads to better SDR and SAR results for the DNN-based technique, while the first setting allows for more interference rejection, and a higher SAR for the KAM method.

Finally, given those results, one should choose a method that is adapted to the target application. If the main goal of the separation is to reduce the overall artifacts and distortion, one should use MaD TwinNet with the baseline mixture's phase in setting 2 (in addition, it is faster than setting 1). If one wants to specifically reduce the artifacts in the harmonic track, then the KAM method is a suitable choice. Finally, if the goal is to reduce interference, it is preferable to use MaD TwinNet with PU-HPSS in the first setting.

\vspace{-6pt}
\section{Conclusion}\label{sec:conclu}\vspace{-6pt}
In this work, we proposed a system for harmonic/percussive source separation based on the MaD TwinNet architecture and further improved with a phase recovery iterative algorithm. This system has demonstrated significant improvement over the baseline KAM. Indeed, MaD TwinNet is useful for reducing the overall distortion compared to KAM, and using a phase recovery algorithm which exploits a sinusoidal model reduces interference in the estimates. Future work will focus on analyzing the filters learned by Mad TwinNet, as this architecture could be more optimally tuned for this specific task. Another interesting future research direction is the joint estimation of magnitude and phase in a unified framework, rather than in a two-stage approach. For instance, a Bayesian framework inspired from~\cite{Nugraha2016,Magron2018a} has a great potential for tackling this issue.

\section{Acknowledgments}
This research was partly funded by the European Research Council under the European Union's H2020 Framework Programme through ERC Grant Agreement 637422 EVERYSOUND. P. Magron is supported by the Academy of Finland, project no.\ 290190. S.-I. Mimilakis is supported by the H2020 Framework Programme (H2020-MSCA-ITN-2014) under grant agreement no 642685 MacSeNet. P. Magron, K. Drossos, and T. Virtanen wish to acknowledge CSC-IT Center for Science, Finland, for computational resources. Part of the computations leading to these results was performed on a TITAN-X GPU donated by NVIDIA to K. Drossos.

\bibliographystyle{IEEEbib}
\bibliography{refs}

\begin{thebibliography}{10}

\bibitem{Comon2010}
P.~Comon and C.~Jutten,
\newblock {\em Handbook of blind source separation: independent component
  analysis and applications},
\newblock Academic press, 2010.

\bibitem{Ono2008}
N.~Ono, K.~Miyamoto, J.~Le Roux, H.~Kameoka, and S.~Sagayama,
\newblock ``Separation of a monaural audio signal into harmonic/percussive
  components by complementary diffusion on spectrogram,''
\newblock in {\em Proc. of European Signal Processing Conference (EUSIPCO)},
  August 2008.

\bibitem{Gkiokas2012}
A.~Gkiokas, V.~Katsouros, G.~Carayannis, and T.~Stajylakis,
\newblock ``Music tempo estimation and beat tracking by applying source
  separation and metrical relations,''
\newblock in {\em Proc. of IEEE International Conference on Acoustics, Speech
  and Signal Processing (ICASSP)}, March 2012.

\bibitem{Driedger2014}
J.~Driedger, M.~M{\"u}ller, and S.~Ewert,
\newblock ``Improving time-scale modification of music signals using
  harmonic-percussive separation,''
\newblock {\em IEEE Signal Processing Letters}, vol. 21, no. 1, pp. 105--109,
  January 2014.

\bibitem{FitzGerald2010}
D.~FitzGerald,
\newblock ``Harmonic/percussive separation using median filtering,''
\newblock in {\em Proc. of International Conference on Digital Audio Effects
  (DAFx)}, September 2010.

\bibitem{FitzGerald2013}
D.~FitzGerald, A.~Liukus, Z.~Rafii, B.~Pardo, and L.~Daudet,
\newblock ``Harmonic/percussive separation using kernel additive modelling,''
\newblock in {\em Proc. of IET Irish Signals Systems Conference}, June 2013.

\bibitem{Liutkus2014}
A.~Liutkus, D.~Fitzgerald, Z.~Rafii, B.~Pardo, and L.~Daudet,
\newblock ``Kernel additive models for source separation,''
\newblock {\em IEEE Transactions on Signal Processing}, vol. 62, no. 16, pp.
  4298--4310, August 2014.

\bibitem{Tachibana2014}
H.~Tachibana, N.~Ono, H.~Kameoka, and S.~Sagayama,
\newblock ``Harmonic/percussive sound separation based on anisotropic
  smoothness of spectrograms,''
\newblock {\em IEEE/ACM Transactions on Audio, Speech, and Language
  Processing}, vol. 22, no. 12, pp. 2059--2073, December 2014.

\bibitem{Cano2014}
E.~Cano, M.~Plumbley, and C.~Dittmar,
\newblock ``{Phase-based harmonic percussive separation},''
\newblock in {\em Proceedings of the Annual Conference of the International
  Speech Communication Association (Interspeech)}, Sept. 2014, pp. 1628--1632.

\bibitem{Lee1999}
Daniel~D. Lee and H.~Sebastian Seung,
\newblock ``{Learning the parts of objects by non-negative matrix
  factorization},''
\newblock {\em Nature}, vol. 401, no. 6755, pp. 788--791, 1999.

\bibitem{Canadas2014}
F.~J. Canadas-Quesada, P.~Vera-Candeas, N.~Ruiz-Reyes, J.~Carabias-Orti, and
  P.~Cabanas-Molero,
\newblock ``Percussive/harmonic sound separation by non-negative matrix
  factorization with smoothness/sparseness constraints,''
\newblock {\em EURASIP Journal on Audio, Speech, and Music Processing}, vol.
  2014, no. 1, July 2014.

\bibitem{Laroche2015}
C.~Laroche, M.~Kowalski, H.~Papadopoulos, and G.~Richard,
\newblock ``A structured nonnegative matrix factorization for source
  separation,''
\newblock in {\em Proc. of European Signal Processing Conference (EUSIPCO)},
  August 2015.

\bibitem{Smaragdis2004}
P.~Smaragdis,
\newblock ``Non-negative matrix factor deconvolution; extraction of multiple
  sound sources from monophonic inputs,''
\newblock in {\em Proc. of International Conference on Independent Component
  Analysis and Signal Separation}, C.~G. Puntonet and A.~Prieto, Eds., 2004.

\bibitem{Laroche2017}
C.~Laroche, H.~Papadopoulos, M.~Kowalski, and G.~Richard,
\newblock ``Drum extraction in single channel audio signals using multi-layer
  non negative matrix factor deconvolution,''
\newblock in {\em Proc. of IEEE International Conference on Acoustics, Speech
  and Signal Processing (ICASSP)}, March 2017.

\bibitem{Nugraha2016}
A.~A. Nugraha, A.~Liutkus, and E.~Vincent,
\newblock ``Multichannel audio source separation with deep neural networks,''
\newblock {\em IEEE/ACM Transactions on Audio, Speech, and Language
  Processing}, vol. 24, no. 9, pp. 1652--1664, September 2016.

\bibitem{Takahashi2017}
N.~Takahashi and Y.~Mitsufuji,
\newblock ``Multi-scale multi-band {DenseNets} for audio source separation,''
\newblock in {\em {Proc. of IEEE Workshop on Applications of Signal Processing
  to Audio and Acoustics (WASPAA)}}, October 2017.

\bibitem{Mimilakis2017}
S.~I. Mimilakis, K.~Drossos, T.~Virtanen, and G.~Schuller,
\newblock ``A recurrent encoder-decoder approach with skip-filtering
  connections for monaural singing voice separation,''
\newblock in {\em Proc. IEEE International Workshop on Machine Learning for
  Signal Processing (MLSP)}, September 2017.

\bibitem{Mimilakis2018}
S.~I. Mimilakis, K.~Drossos, J.~F. Santos, G.~Schuller, T.~Virtanen, and
  Y.~Bengio,
\newblock ``Monaural singing voice separation with skip-filtering connections
  and recurrent inference of time-frequency mask,''
\newblock in {\em {Proc. IEEE International Conference on Acoustics, Speech and
  Signal Processing (ICASSP)}}, April 2018.

\bibitem{Drossos2018}
K.~Drossos, S.~I. Mimilakis, D.~Serdyuk, G.~Schuller, T.~Virtanen, and
  Y.~Bengio,
\newblock ``{MaD TwinNet}: Masker-denoiser architecture with twin networks for
  monaural sound source separation,''
\newblock in {\em {Proc. IEEE International Joint Conference on Neural Networks
  (IJCNN)}}, July 2018.

\bibitem{Lim2017}
W.~Lim and T.~Lee,
\newblock ``Harmonic and percussive source separation using a convolutional
  auto encoder,''
\newblock in {\em Proc. European Signal Processing Conference (EUSIPCO)},
  August 2017.

\bibitem{Magron2018}
P.~Magron, R.~Badeau, and B.~David,
\newblock ``Model-based {STFT} phase recovery for audio source separation,''
\newblock {\em {IEEE/ACM Transactions on Audio, Speech and Language
  Processing}}, vol. 26, no. 6, pp. 1095--1105, June 2018.

\bibitem{Liutkus2017}
A.~Liutkus, F.-R. St{\"o}ter, Z.~Rafii, D.~Kitamura, B.~Rivet, N.~Ito, N.~Ono,
  and J.~Fontecave,
\newblock ``{The 2016 Signal Separation Evaluation Campaign},''
\newblock in {\em {Proc. International Conference on Latent Variable Analysis
  and Signal Separation (LVA/ICA)}}, February 2017.

\bibitem{Serdyuk2018}
D.~Serdyuk, N.-R. Ke, A.~Sordoni, A.~Trischler, C.~Pal, and Y.~Bengio,
\newblock ``{Twin Networks}: Matching the future for sequence generation,''
\newblock in {\em {Proc. of International Conference on Learning
  Representations (ICLR)}}, April 2018.

\bibitem{vincent:2010:jml}
P.~Vincent, H.~Larochelle, I.~Lajoie, Y.~Bengio, and P.-A. Manzagol,
\newblock ``Stacked denoising autoencoders: Learning useful representations in
  a deep network with a local denoising criterion,''
\newblock {\em Journal of Machine Learning Research}, vol. 11, pp. 3371--3408,
  2010.

\bibitem{bengio:1994:tnp}
Y.~Bengio, P.~Simard, and P.~Frasconi,
\newblock ``Learning long-term dependencies with gradient descent is
  difficult,''
\newblock {\em IEEE Transactions on Neural Networks}, vol. 5, no. 2, pp.
  157--166, Mar 1994.

\bibitem{hochreiter:1997:lsm}
S.~Hochreiter and J.~Schmidhuber,
\newblock ``Long short-term memory,''
\newblock {\em Neural Comput.}, vol. 9, no. 8, pp. 1735--1780, Nov. 1997.

\bibitem{grave:2016:arxiv}
E.~Grave, A.~Joulin, and N.~Usunier,
\newblock ``Improving neural language models with a continuous cache,''
\newblock {\em CoRR}, vol. abs/1612.04426, 2016.

\bibitem{gulcehre:2017:arxiv}
{\c{C}}.~G{\"{u}}l{\c{c}}ehre, S.~Chandar, and Y.~Bengio,
\newblock ``Memory augmented neural networks with wormhole connections,''
\newblock {\em CoRR}, vol. abs/1701.08718, 2017.

\bibitem{dieng:2016:arxiv}
A.~B. Dieng, C.~Wang, J.~Gao, and J.~W. Paisley,
\newblock ``Topicrnn: {A} recurrent neural network with long-range semantic
  dependency,''
\newblock {\em CoRR}, vol. abs/1611.01702, 2016.

\bibitem{Krawczyk2014}
M.~Krawczyk and T.~Gerkmann,
\newblock ``{STFT} phase reconstruction in voiced speech for an improved
  single-channel speech enhancement,''
\newblock {\em IEEE/ACM Transactions on Audio, Speech, and Language
  Processing}, vol. 22, no. 12, pp. 1931--1940, December 2014.

\bibitem{Abe2004}
M.~Abe and J.~O. Smith,
\newblock ``{Design criteria for simple sinusoidal parameter estimation based
  on quadratic interpolation of {FFT} magnitude peaks},''
\newblock in {\em {Audio Engineering Society Convention 117}}, May 2004.

\bibitem{Vincent2006}
E.~Vincent, R.~Gribonval, and C.~F{\'e}votte,
\newblock ``Performance measurement in blind audio source separation,''
\newblock {\em IEEE Transactions on Speech and Audio Processing}, vol. 14, no.
  4, pp. 1462--1469, July 2006.

\bibitem{Raffel2014}
C.~Raffel, B.~McFee, E.~J. Humphrey, J.~Salamon, O.~Nieto, D.~Liang, and
  D.~P.~W. Ellis,
\newblock ``{mir}\_{eval}: A transparent implementation of common {MIR}
  metrics,''
\newblock in {\em Proc. International Society for Music Information Retrieval
  Conference (ISMIR)}, October 2014.

\bibitem{Magron2018c}
P.~Magron, K.~Drossos, S.~I. Mimilakis, and T.~Virtanen,
\newblock ``Reducing interference with phase recovery in {DNN}-based monaural
  singing voice separation,''
\newblock in {\em {Proc. of Interspeech}}, September 2018.

\bibitem{Magron2018a}
P.~Magron and T.~Virtanen,
\newblock ``{Bayesian anisotropic Gaussian model for audio source
  separation},''
\newblock in {\em {Proc. of IEEE International Conference on Acoustics, Speech
  and Signal Processing (ICASSP)}}, April 2018.

\end{thebibliography}

\end{document}